\documentclass[final,1p,times]{elsarticle} 
\usepackage{graphicx} 
\usepackage{amssymb} 
\usepackage{amsthm} 
\usepackage{lineno} 
 \def\BA{\begin{eqnarray}}
 \def\BE{\begin{equation}}
 \def\BF{\begin{figure}[htb]}
 \def\BT{\begin{table}[htb]}
 \def\EA{\end{eqnarray}}
 \def\EE{\end{equation}}
 \def\EF{\end{figure}}
 \def\ET{\end{table}}

 \def\lsim{\mathrel{\rlap{\lower4pt\hbox{\hskip1pt$\sim$}}
     \raise1pt\hbox{$<$}}}         
 \def\gsim{\mathrel{\rlap{\lower4pt\hbox{\hskip1pt$\sim$}}
     \raise1pt\hbox{$>$}}}         

\journal{Nuclear Physics A} 
\begin{document} 

\begin{frontmatter} 


\title{Physics of Large-x Nuclear Suppression}

\author{J.~Nemchik$^{a,b}$ and M.~\v Sumbera$^{c}$}

\address[a]{Institute of Experimental Physics SAS,
Watsonova 47, 04001 Ko\v sice, Slovakia}
\address[b]{Czech Tech. Univ. in Prague, FNSPE,
B\v rehov\' a 7, 11519 Prague, Czech Republic}
\address[c]{Nuclear Physics Institute AS CR,
25068 \v Re\v z/Prague, Czech Republic}
%

\begin{abstract}
We discuss a common feature of all known reactions on nuclear
targets - a significant suppression at large $x$. Simple
interpretation of this effect is based on energy conservation
restrictions in initial state parton rescatterings. Using the
light-cone dipole approach this mechanism is shown to control
variety of processes on nuclear targets: high-$p_T$ particle
production at different rapidities as well as direct and virtual
(Drell-Yan) photon production. We demonstrate universality and
wide applicability of this mechanism allowing to describe
large-$x$ effects also at SPS and FNAL energies too low for the
onset of coherent effects or shadowing.
\end{abstract}

\end{frontmatter} 




%
%
%
\section{Introduction}
\vspace*{-.20cm}

Recent measurements of high-$p_T$ hadrons produced in the beam
fragmentation region in $d+Au$ collisions at RHIC
\cite{brahms,brahms-07,star} allow to reach the smallest values of
Bjorken $x$ in the nucleus and thus to reach maximal coherence
effects leading eventually to nuclear suppression. Observed
suppression is usually interpreted within the models based on
color glass condensate (CGC). However, such an interpretation
misses a global applicability. For example, a similar suppression
like at RHIC was measured also in $p+Pb$ collisions at SPS where
no effects of coherence are possible.

The rise of the suppression  with Feynman $x_F$  for hadrons
produced in $p+Pb$ collisions at SPS \cite{na49}  or for the
Drell-Yan (DY) pairs at FNAL \cite{e772} has a similar pattern as
seen at RHIC. All these examples and another reactions treated in
\cite{knpsj-05} favor the same large-$x$ mechanism independently
of the energy and type of reaction.

Such a common mechanism was proposed in \cite{knpsj-05,npps-08}
where large-$x_F$ nuclear suppression was shown to be caused by
the energy conservation in multiple parton rescatterings. This
mechanism is a leading twist effect giving rise to the breakdown
of QCD factorization and exhibits also the $x_F$-scaling
\cite{npps-08}.

Another consequence of this treatment discussed in this paper is
the manifestation of nuclear effects also at midrapidities, i.e.
at large $x_T = 2\,p_T/\sqrt{s}$. We expect a suppression pattern
similar to that at large $x_F$ and the nucleus-to-nucleon ratio
below one. Similarly to $x_F$-scaling at forward rapidities
$x_T$-scaling of this effect is predicted.

In this paper we further exploit proposed model
\cite{knpsj-05,npps-08} to analyze and quantify the nuclear
suppression at large $x$ for a variety of processes occurring in
$p(d)+A$ and $A+B$ collisions.

\vspace*{-.30cm}

%
%
\section{Sudakov suppression, production cross section
\label{sudakov}}
\vspace*{-.20cm}

In any hard reaction in the limit $x\to 1$ gluon radiation is
forbidden by the energy conservation. Then the probability to have
a large rapidity gap (LRG) $\Delta y = -\ln S(x)$ between the
leading parton and rest of the system, where the Sudakov
suppression factor $S(x) = 1-x$ \cite{knpsj-05}.

Suppression at $x\to 1$ can thus be formulated in terms of
multiple interactions of projectile partons with the nucleus. Each
of multiple interactions produces an extra suppression factor
$S(x)$ and corresponding weight factors are given by the AGK
cutting rules \cite{agk}. Then, in terms of the nuclear thickness
function $T_A(b)$ and the effective cross section $\sigma_{eff}$
\cite{knpsj-05}, the cross sections of hard reaction on a nuclear
target $A$ at impact parameter $b$ and on a nucleon $N$ are
related as \cite{prepar1}, \vspace*{-0.2cm}
%
%
 \BA
\frac{d~^2\sigma_A}{dx d~^2b} =
\frac{d\sigma_N}{dx} \frac{1}{\sigma_{eff}}
e^{- \sigma_{eff} T_A(b)}
\sum\limits_{n=1}^A\,\frac{n}{n!}\,\left[\sigma_{eff}\,T_A(b)
\right]^n S(x)^{n-1} =
\frac{d\sigma_N}{dx} T_A(b) e^{- [1-S(x)] \sigma_{eff} T_A(b)}\, .
 \label{70}
 \EA
%
%

\vspace*{-0.2cm}
Employing factorization,
the hadron production cross section
in $d+A\,(p+p)$ collisions reads,

\vspace*{-0.3cm}
%
%
 \BA
\frac{d~^2\sigma(p~A\rightarrow h~X)}{d~^2p_T\,d\eta} =
\int\limits_{x_h}^{1} dx\,
\int\limits_{z_{min}}^1 \frac{dz}{z~^2}\,
\Biggl\{
\sum\limits_q\, f_{q/d(p)}(x,q_T^2)\,
\Biggl [
\frac{d~^2\sigma(q~A(N)\rightarrow q~X)}{d~^2q_T\,d\eta}\,
D_{h/q}(z,q_T^2) +
\nonumber\\
\frac{d~^2\sigma(q~A(N)\rightarrow g~X)}{d~^2q_T\,d\eta}\,
D_{h/g}(z,q_T^2) \Biggr ] +
f_{g/d(p)}(x,q_T^2)\,
\frac{d~^2\sigma(g~A(N)\rightarrow g~X)}{d~^2q_T\,d\eta}\,
D_{h/g}(z,q_T^2) \Biggr\}\, ,
\label{80}
 \EA
%
%

\vspace*{-0.2cm} \hspace*{-0.63cm} where $\eta$ is pseudorapidity,
$x_h = p_T\,e^{\eta}/\sqrt{s}$, $z_{min} = x_h/x$ and $q_T =
p_T/z$ is the parton transverse momentum. The parton cross
sections $d^{2}\sigma(q(g)+A(N))/d^2q_T d\eta$ in Eq.~(\ref{80}) are
calculated in the light-cone (LC) dipole approach
\cite{knpsj-05,prepar1,knst-01}. For parton distribution functions
we use the parametrization from \cite{grv}. Fragmentation
functions were taken from \cite{fs-07}.

As first shown in \cite{knpsj-05,npps-08} the effective projectile
quark (gluon) distribution correlates with the target
and corresponding quark (gluon) distribution in the nucleus reads
\vspace*{-0.2cm}
%
%
 \BE
f^{(A)}_{q(g)/N}\bigl (x,q_T^2\bigr ) =
C\,f_{q(g)/N}\bigl (x,q_T^2\bigr )\, \frac{\int d~^2b\,
\left[e^{-x\,\sigma_{eff}T_A(b)}- e^{-\sigma_{eff}T_A(b)}\right]}
{(1-x)\int d~^2b\,\left[1- e^{-\sigma_{eff}T_A(b)}\right]}\, .
\label{100}
 \EE
%
%
For the quark part
the normalization factor $C$ in
Eq.~(\ref{100}) is fixed by the Gottfried sum rule.

\vspace*{-.30cm}

%
%
\section{Hadron production at forward rapidities}
\vspace*{-.20cm}

In 2004 the BRAHMS Collaboration \cite{brahms} observed a
significant nuclear suppression of $h^-$ at $\eta=3.2$. Much
stronger nuclear effects was found later on by the STAR
Collaboration \cite{star} for $\pi^0$ production at $\eta = 4.0$.
All these data are consistent with model calculations
\cite{knpsj-05,npps-08}. A strong rise of suppression with $\eta$
reflects much smaller survival probability $S(x)$ of a LRG at
larger $x$.

Since parton energy loss is proportional to the initial energy the
energy conservation restrictions in  multiple parton rescatterings
should also lead to $x$-scaling of the nuclear effects . A
similarity of suppression at different energies and
pseudorapidities was demonstrated in \cite{knpsj-05,npps-08}.
\vspace*{-.30cm}

%
%
\section{Hadron production at midrapidities}
\vspace*{-.20cm}

\begin{figure}[tbh]
\hspace*{0.65cm}

\includegraphics[scale=0.34]{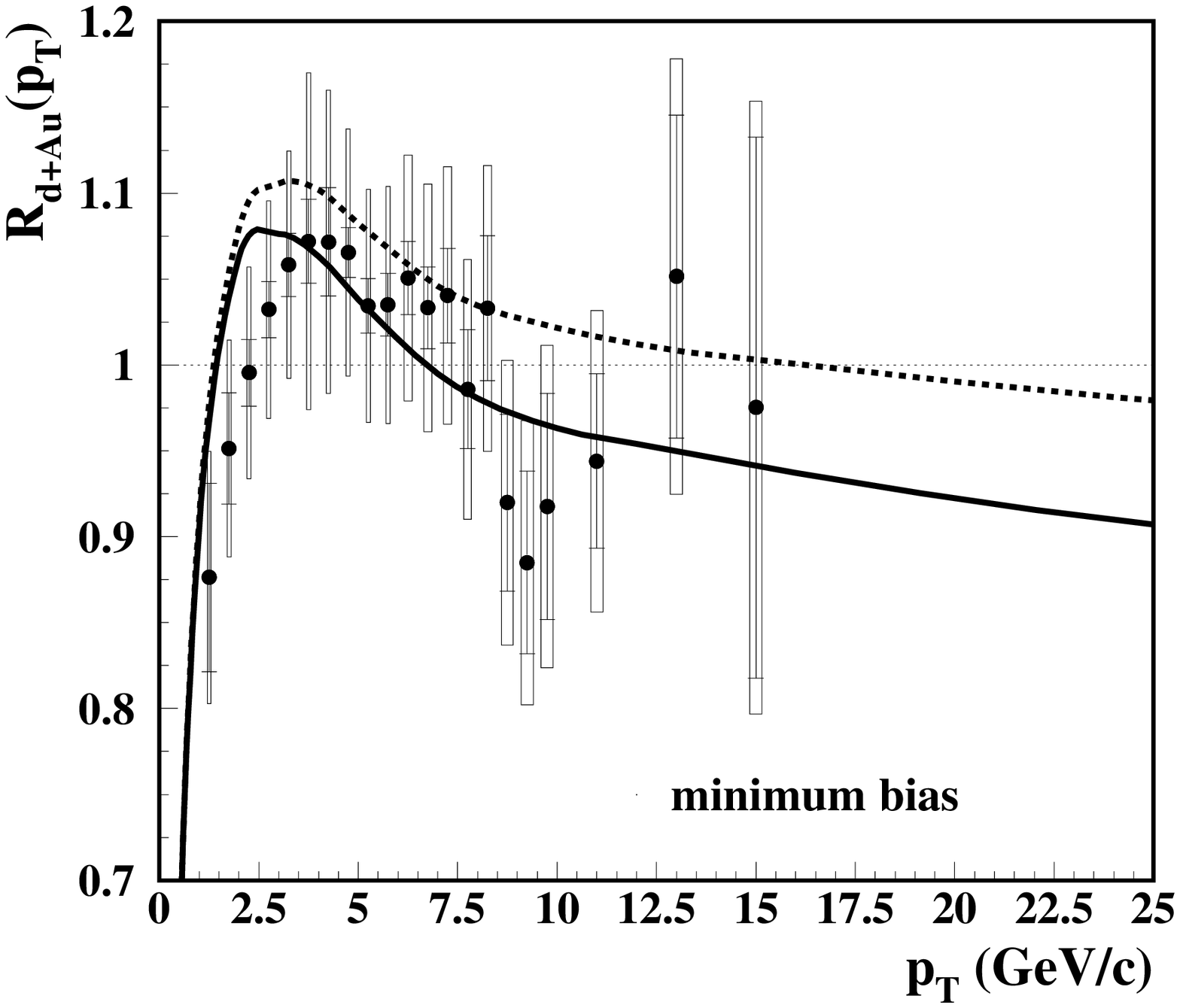}
\hspace*{.95cm}
\includegraphics[scale=0.34]{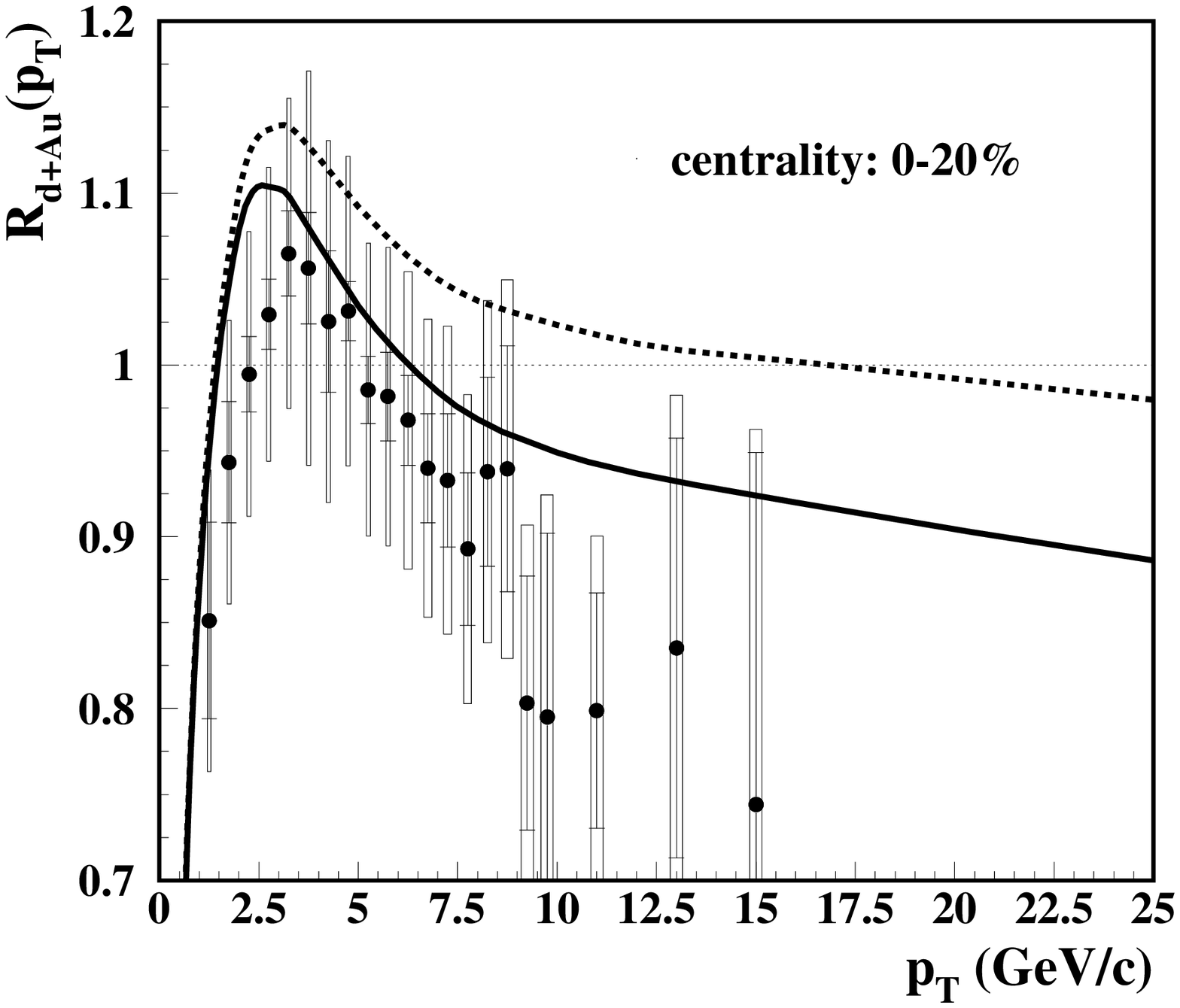}
\begin{center}

\vspace*{-.5cm}
\caption
{
     Nuclear modification factor for $\pi^0$ produced in $d+Au$
     collisions at
$\sqrt{s}=200$\,GeV corresponding to minimum bias (Left) and
centrality 0-20\% (Right), respectively. Solid and dashed lines
represent calculations with and without energy conservation
in multiple parton rescatterings, respectively.
The data are from the
PHENIX Collaboration \cite{phenix}
}
\label{midrap}
\end{center}
\vspace*{-1.2cm}

\end{figure}

Another manifestation of the energy conservation in multiple
parton rescatterings occurs at midrapidities. Here the
corresponding values of $p_T$ should be high enough to keep $x_T =
2\,p_T/\sqrt{s}$ on the same level as $x_F$ at the forward
rapidities. This is supported by data from the PHENIX
Collaboration \cite{phenix} showing an evidence for suppression at
large $p_T\gsim 8\,$GeV/c (see Fig.~\ref{midrap}).
%

If effects of energy conservation are
not included the $p_T$ dependence of
$R_{d+Au}$ described by the
dashed lines exhibits only
a small suppression at large $p_T$
given by the isotopic effects
(see Fig.~\ref{midrap}).
After inclusion of energy sharing in
parton rescatterings we predict $R_{d+Au} < 1$
at large $p_T$
presented by the solid lines.
More precise data are needed
for a clear manifestation of breakdown of QCD factorization.

%
%
\section{Direct photon production in Au+Au collisions at RHIC}
\label{dp}
\vspace*{-.2cm}

\begin{figure}[htb]

\includegraphics[scale=0.35]{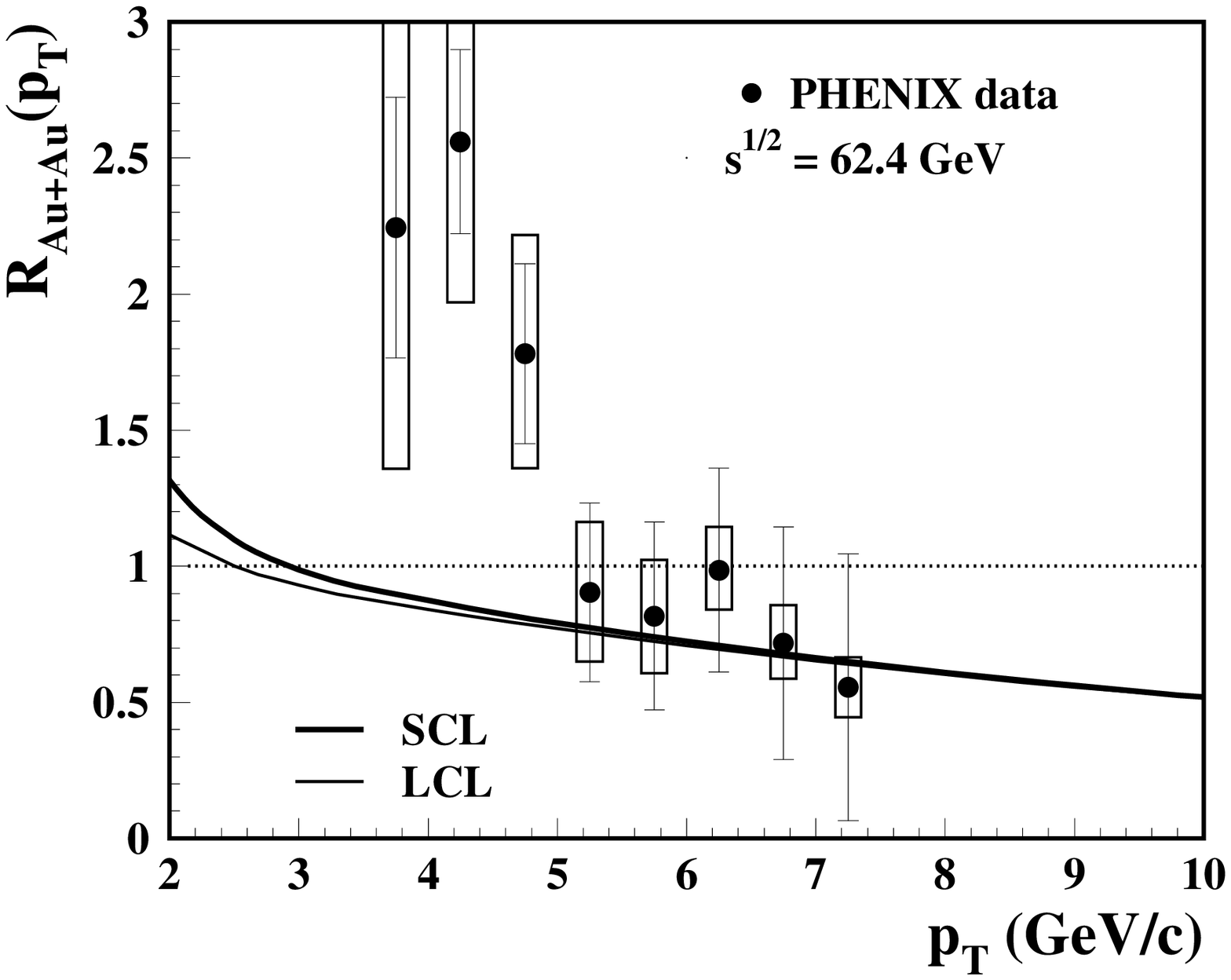}
\hspace*{0.85cm}
\includegraphics[scale=0.35]{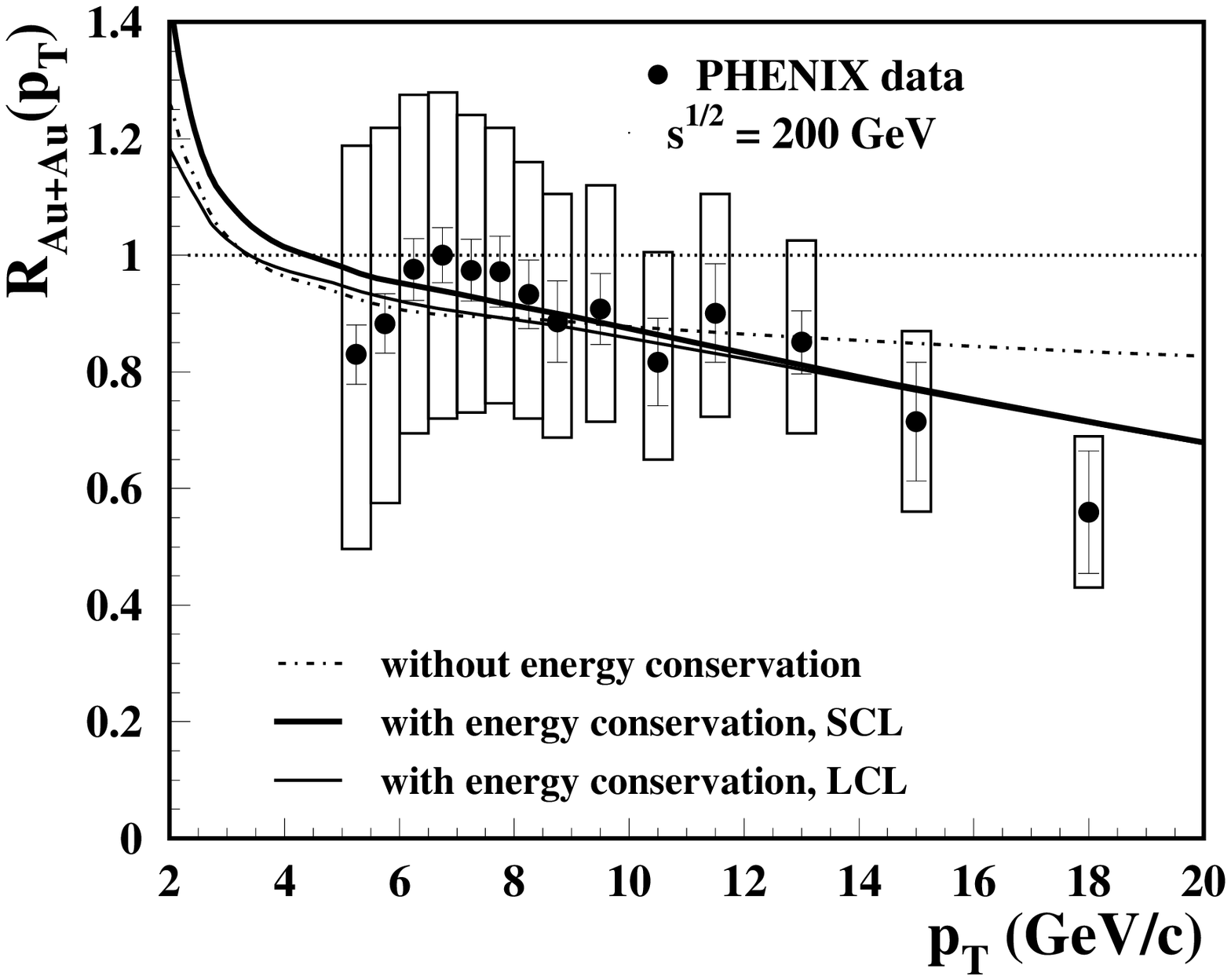}
\begin{center}

\vspace*{-.5cm}
\caption
{
Nuclear modification factor for direct photon production in $Au+Au$
collisions at
$\sqrt{s}=62\,$GeV (Left) and $\sqrt{s}=200\,$GeV (Right)
vs. PHENIX data \cite{phenix}.
Dashed line represents calculations without energy conservation.
Thick and thin
solid lines represent calculations in the limit of long
(LCL) and short (SCL) coherence length, respectively.
}
\label{photons}
\end{center}
\vspace*{-1.3cm}

\end{figure}

Direct photons in $Au+Au$ collisions are also suppressed at large
$p_T$ as was demonstrated by the PHENIX
Collaboration~\cite{phenix-dp}. Model predictions for the ratio
$R_{Au+Au}$ as a function of $p_T$ are compared with data in
Fig.~\ref{photons}. Expressions for production cross sections have
been adopted from ~\cite{kst1,prepar2}. If the energy conservation
in parton rescatterings is not taken into account model
calculations depicted by the dashed line gives a value
$R_{Au+Au}\rightarrow 0.8$ in accord with onset of isotopic
effects. Inclusion of the energy conservation leads to strong
nuclear effects at large $p_T$ as is demonstrated by thick and
thin solid lines.

\vspace*{-.2cm}
%
%
\section{Nuclear suppression at SPS and FNAL energies}
\label{na49}
\vspace*{-.3cm}

\begin{figure}[hbt]
\vspace*{-.8cm}
\hspace*{-0.35cm}

\includegraphics[width=0.43\textwidth]{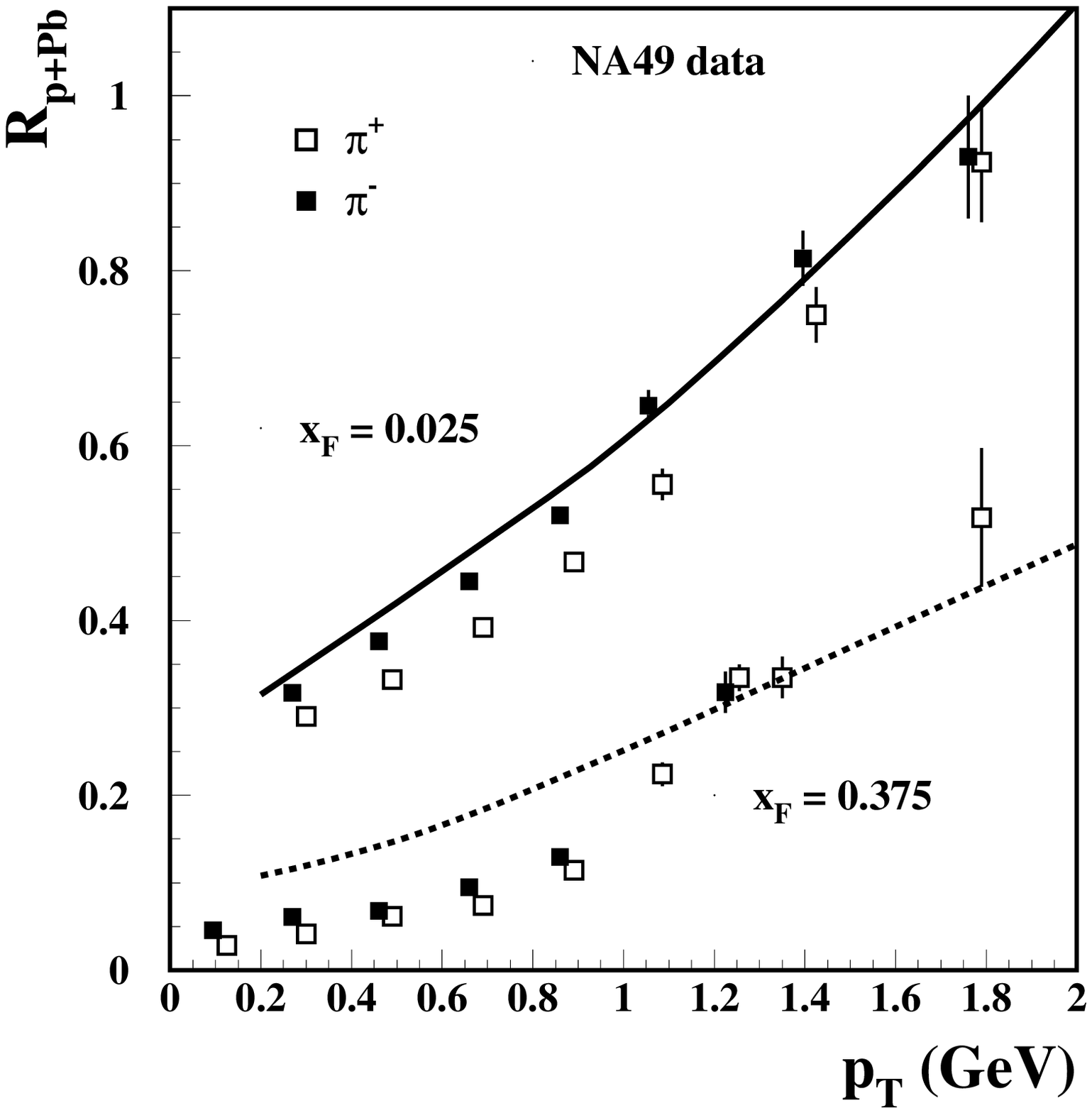}
\hspace*{0.45cm}
\includegraphics[width=0.51\textwidth]{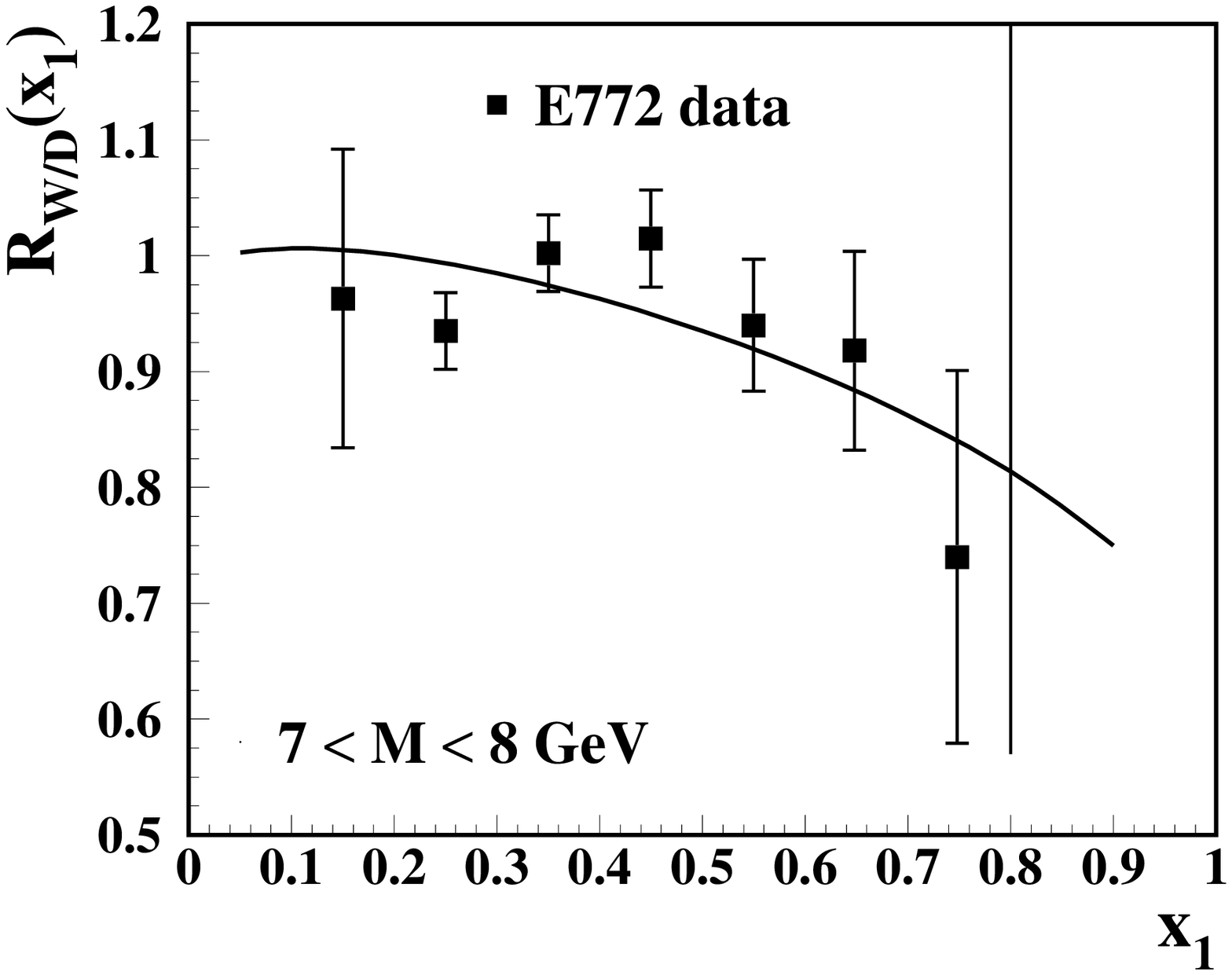}
\begin{center}

\vspace*{-.6cm}
\caption
{
(Left)
Ratio, $R_{p+Pb}(p_T)$, for $\pi^\pm$ production rates
in $p+Pb$ and $p+p$ collisions as function of $p_T$ at two fixed
$x_F = 0.025$ and $0.375$ vs. NA49 data
\cite{na49}.
(Right)
Normalized ratio of Drell-Yan cross sections
on Tungsten and Deuterium as a function of $x_1$
vs. E772 data \cite{e772}.
}
\label{small}
\end{center}
\vspace*{-1.1cm}

 \end{figure}

The left panel of Fig.~\ref{small} clearly manifests that pions
from $p+Pb$ collisions at SPS energy exhibit the same suppression
pattern as that in the RHIC kinematic range. Model predictions
employ the dipole formalism for calculation of nuclear broadening
using standard convolution expression based on QCD factorization
\cite{knst-01}. Initial state multiple interactions leading to
breakdown of QCD factorization are included as described in
Sect.~\ref{sudakov}. One can see a reasonable agreement of our
calculations with NA49 data \cite{na49}.
%

The DY reaction is also known to be considerably suppressed at
large $x_F$ ($x_1$) \cite{johnson} (see the right panel of
Fig.~\ref{small}). Using the same mechanism as discussed in
Sect.~\ref{sudakov} one can explain a strong suppression at large
$x_1$. The differential cross section for the photon radiation in
a quark-nucleus collision is calculated~\cite{prepar3} using the
LC Green function formalism~\cite{kst1}. The right panel of
Fig.~\ref{small} demonstrates a good agreement of our calculations
with E772 data~\cite{e772}. \vspace*{-0.5cm}

%
%
\section{Summary}\label{conclusions}
\vspace*{-.2cm}

Unified approach to large-$x$ nuclear suppression based on the
energy conservation restrictions in multiple parton rescatterings
was presented.
QCD factorization fails at the kinematic
limit, $x \rightarrow 1$.
Universal
suppression driven by Sudakov factor $S(x)$ brings in the
$x$-scaling of nuclear effects.
The same formalism explains well available data from RHIC on
suppression of high-$p_T$ hadrons and photons at different
rapidities.
This common mechanism explains also
a suppression at low SPS and FNAL
energies where no coherence effects are possible.
\\
\vspace*{-0.2cm}

{\bf Acknowledgements :} This work was supported in part by the
Grant Agency of the Czech Republic, Grant 202/07/0079, by the
Slovak Funding Agency, Grant 2/7058/27 and by Grants VZ M\v{S}MT
6840770039 and LC 07048 (Ministry of Education of the Czech
Republic).  
\end{document}